\documentstyle[12pt]{article}
\author{
Serge Galam\\ \\ \\ 
Laboratoire des Milieux D\'{e}sordonn\'{e}s et H\'{e}t\'{e}rog\`{e}nes
\footnotemark[1]\\
Tour 13 - Case 86, 4 place Jussieu, \\ 75252 Paris Cedex 05, France\\}
\title{Comment on ``A landscape theory of aggregation"} 
\addtocounter{footnote}{0}
\footnotetext{Laboratoire associ\'{e} au CNRS (URA n$^{\circ}$ 800) et \`{a} l'Universit\'{e} 
P. et M. Curie - Paris 6}
\date{B. J. Pol. S. \underline{28} (1998) 411-412}
\addtocounter{footnote}{1}
\begin{document}
\maketitle
\baselineskip 3.3ex
\footskip 5ex
\parindent 2.5em
\abovedisplayskip 5ex
\belowdisplayskip 5ex
\abovedisplayshortskip 3ex
\belowdisplayshortskip 5ex
\textfloatsep 7ex
\intextsep 7ex

%%%%%%%%%%%%%%%%%%%%%%%%%%%%%%%%%%%%%%%%%%%%%%%%%%%%%%%%%%%%%%%%%%%%%%%
\begin{abstract}
The problem of aggregation processes in alignments is the subject of a paper published
recently in a Statistical Physics Journal (Physica A\underline{230}, 174-188, 1996). 
Two models are presented and discussed in that paper.
First the energy landscape model proposed by Axelrod and Bennett (B. J. Pol. S. \underline{23}, 211-233,
1993), is analysed. 
The model is shown not to
include most of its claimed results. Then a second model is presented to reformulate correctly the problem  
within statistical physics and to extend it beyond the initial Axelrod-Bennett analogy.  

\end{abstract}
\newpage
%%%%%%%%%%%%%%%%%%%%%%%%%%%%%%%%%%%%%%%%%%%%%%%%%%%%%%%%%%%%%%%%%%%%%%% 

Mathematical tools and physical concepts might be a promising way to
describe social collective
phenomena. Several attempts along these lines have
been made, in particular to study 
political organisations [1], voting systems [2], and group
decision making [3].
However, such an approach should be carefully controlled. A straightforward
mapping
of a physical theory built
for a physical reality onto a social reality could be rather misleading.

In their work Axelrod and Bennett (AB) used the physical concept of minimum
energy to build a
landscape model of aggregation [4]. On this basis, they study the 
coalitions which countries or firms could make to optimize their respective relationship, which is
certainly an interesting problem. To achieve their purpose, they constructed a model of magnetic disorder 
from the
available data for propensities of countries or firms to co-operate or to conflict.
Using their model, they drewn several conclusions based on the existence of local frustration between 
the interacting parties [5].

However, there was some confusion in their use of physics, and they did not stick to their equations.
In their model, unfortunately, the disorder is only apparent in the existence of just two energy
minima.
It is called the Mattis spin glass model [5]. It has been shown that performing an appropriate change of 
variables, removes the disorder and the model then becomes identical to a well ordered system, 
the zero temperature
finite size ferromagnetic Ising model [6].

In contrast, most AB comments and conclusions are based, on the existence of frustration in
the countries or firms interactions [5]. Such local frustration would produce a degeneracy of
the energy landscape which in turn would yield instabilities in the global system. 
However, there is no frustration in the model they derived from their data.

In fact they are confusing two models associated with disordered magnetic systems: one without frustration,
the Mattis spin glass model,
and one with frustration, the Edwards-Anderson spin glass model [5].
The AB model turns out
to be of the Mattis spin glass type, while all their comments are drawn from
the physics associated with an Edwards-Anderson spin glass model. Most of Axelrod and Bennett's
conclusions cannot be drawn from their model. 

To demonstrate our statement requires the use of some mathematical technicalities 
which are lenghty and not appropriate to the present journal. Therefore our 
demonstration has been published in a Physics journal [7], where first, 
the AB model is analysed within the field of
Statistical Physics [6] and then the conclusions mentioned above are demonstrated.  
Furthermore, we are able to build up a new coalition model to
describe alignment and competition among a group of actors [7].
Our model does embody the main properties claimed in the AB model. 
Morevover it also predicts new behavior related to the dynamics of bimodal coalitions.
In particular the stability of the cold 
war period and the East European fragmentation process induced by the collapse of the 
Warsaw pact are given an explanation. 
%%%%%%%%%%%%%%%%%%%%%%%%%%%%%%

%%%%%%%%%%%%%%%%%%%%
\end{document}